\documentclass[11pt]{article}
\usepackage{graphics}
\usepackage{graphicx}
  \usepackage{amssymb}
  \usepackage{amsmath}
  \usepackage{ amscd}

     \renewcommand{\baselinestretch}{1.4}
  \renewcommand{\arraystretch}{1.2}
\begin{document}

       \title{A Fast String Matching Algorithm Based on \\
       Lowlight Characters in the Pattern
         }
   \author{Zhengjun Cao$^1$, \qquad Lihua Liu$^2$\\
 {\small $^1$Department of Mathematics, Shanghai University,
  China. \textsf{caozhj\,@\,shu.edu.cn}}\\
 {\small $^2$Department of Mathematics, Shanghai Maritime University, Shanghai,
  China.}  }

 \date{}\maketitle

  \begin{abstract} We put forth a new string matching algorithm which matches the pattern from neither the left nor the right end, instead a special position.  Comparing with the Knuth-Morris-Pratt
    algorithm and the Boyer-Moore algorithm, the new algorithm is more flexible to pick the position for starting comparisons. The option really brings it a saving in cost.

 \textbf{Keywords.} string matching; Knuth-Morris-Pratt
    algorithm; Boyer-Moore algorithm;
     finite automata; lowlight character.

     \textbf{MSC}(2010): 68Qxx, 68P10.
\end{abstract}

 \section{ Introduction }

There are various matching problems, such as approximate string matching \cite{AAK09,AEK11}, inverse pattern matching \cite{AAL97}, two-dimensional pattern matching \cite{ABC04}, real scaled matching \cite{ABL99}, scaled dictionary matching \cite{AC00}, property matching \cite{ACI08},  weighted matching \cite{ACI08}, overlap matching \cite{ACH03}, approximate swapped matching \cite{ALP02}, combinatorial pattern matching \cite{AP12} and speculative parallel pattern matching \cite{LSE11}. Among these matching problems, searching for a word in a natural language text is of great importance. It
 is an important utility in text editors and word processers. Typically, the text
is a document being edited, and the pattern searched for is a
particular word supplied by the user.
 There are three general ways of matching
technique, standard matching algorithm, Knuth-Morris-Pratt
    algorithm \cite{KMP77} and Boyer-Moore algorithm \cite{BM77}. All of them are done from the perspective of
character strings. These techniques could, however, be used to
search for any string of bits or bytes in a binary file.

In this paper, we introduce a new string matching algorithm,
more precisely, for character string matching, not for bits.
  It makes use of  that  each English
alphabet has its own statistical probability.  Unlike
  Knuth-Morris-Pratt algorithm which matches the pattern from the left end and Boyer-Moore algorithm which matches from the right end, our algorithm matches  from a special position.
   The statistical  probability of
the character in the position  is the smallest among that of all
characters in the pattern string. We call such a character (may be not unique) a \emph{lowlight character} of the pattern string. We shall compare the algorithm with two popular algorithms for string matching, the Knuth-Morris-Pratt  algorithm and the Boyer-Moore algorithm.
The flexible option to pick the position for starting comparisons really brings the new algorithm a saving in cost.

  \section{Three general algorithms for string matching }

  \subsection{Standard algorithm}

   In the standard algorithm, we begin by comparing the first
  character of the text with the first character of the substring.
  If they match, we move to the next character of each. This
  process continues until the entire substring matches the text or
  the next characters do not match. See the following example for details.

     \begin{center}  \renewcommand{\baselinestretch}{.8}
  \renewcommand{\arraystretch}{1.3} \small
  \begin{tabular}{|l|l|c|}
  \hline
  & & comparisons \\ \hline
Text: & there  they are & \\
 Pass 1:&  they & 4\\
  Text: & there  they are &  \\
   Pass 2:& \ \,they &1\\
    Text:& there  they are & \\
    Pass 3:
  & \ \ \  they &1\\
    Text:& there  they are  &\\
Pass 4:& \ \  \ \ they & 1\\
 Text:& there  they are & \\
 Pass 5:& \ \ \ \ \ \,they & 1\\
Text:& there  they are &\\
Pass 6:& \qquad  \,they &1\\
 Text: & there  they are & \\
Pass 7:&  \ \ \ \ \ \ \ \,they & 4
\\ \hline
  \end{tabular} \end{center}

It is easy to find that the standard algorithm wastes a lot of effort. If we have
matched the beginning part of the substring, we can use that
information to tell us how far to move in the text to start the
next match.

 \subsection{Knuth-Morris-Pratt algorithm}

The Knuth-Morris-Pratt algorithm is based on finite automata but
uses a simpler method of handling the situation of when the
characters don't match. In the Knuth-Morris-Pratt algorithm, we label the states
with the symbol that should match at that point. We then only need
two links from each state, one for a successful match and the
other for a failure. The success link will take us to the next node
in the chain, and the failure link will take us back to a previous
node based on the word pattern. Each success link of  a
Knuth-Morris-Pratt automata causes the ``fetch" of a new character
from the text. Failure links do not get a new character but reuse
the last character fetched. If we reach the final state, we know
that we found the substring.


 \subsection{Boyer-Moore algorithm}

 The Boyer-Moore algorithm is different from the previous two
 algorithms in that it matches the pattern from the right instead
 of left end. For example, in the following example, we first compare the \textit{y}
 with the \textit{r} and find a mismatch. Because \textit{r} doesn't appear in the
 pattern at all, we know the pattern can be moved to the right a
 full four characters (the size of the pattern). We next compare
 the \textit{y} with the \textit{h} and find a mismatch. This time because the \textit{h}
 does appear in the pattern, we move the pattern only two
 characters to the right so that the \textit{h} characters line up. We then
 begin the match from the right side and find a complete match for
 the pattern.

  \begin{center}  \renewcommand{\baselinestretch}{1.0}
  \renewcommand{\arraystretch}{1.3} \small
  \begin{tabular}{|l|l|c|}
  \hline
  & & comparisons \\ \hline
Text: & there  they are & \\
 Pass 1:&  they & 1\\
  Text: & there \ they are &  \\
   Pass 2:& \ \ \ \  \ \,they &1\\
    Text:& there  they are & \\
    Pass 3:
  &  \ \ \ \ \ \ \  \ they &4
 \\ \hline
  \end{tabular} \end{center}

In the Boyer-Moore algorithm,  we did 6 character
 comparisons verses 13 in the standard algorithm.

\section{New algorithm   }

\subsection{Description and examples }

All algorithms mentioned above do not
consider each English alphabet has its own statistical probability.
Whereas the language property is very useful in daily life, especially in cryptanalysis \cite{S05}.

\centerline{Table 1: Statistical probabilities of English alphabets}
\begin{center}
\renewcommand{\baselinestretch}{1.1}
  \renewcommand{\arraystretch}{1.0}
  \small
  \begin{tabular}{|l|l||l|l|}
  \hline
character &   probability  &  character &   probability \\
\hline   A & 0.082 &   N & 0.067 \\
  B& 0.015 &  O& 0.075\\
  C& 0.028 &  P& 0.019\\
  D& 0.042 &  Q& 0.001 \\
  E& 0.127 &  R& 0.060 \\
  F& 0.022 &  S& 0.063 \\
  G& 0.020 &  T& 0.091\\
  H& 0.061 &  U& 0.028 \\
  I& 0.070 &  V& 0.010 \\
  J& 0.002 &  W& 0.023 \\
  K& 0.008 &  X& 0.001 \\
  L& 0.040 &  Y& 0.020 \\
  M& 0.024 &  Z& 0.001
  \\ \hline
  \end{tabular}
   \end{center}

The basic idea behind the new algorithm  is to find a character (may be not unique) which has the smallest probability among that of all characters in the pattern string.  For convenience, we call such a character
 \emph{lowlight character} in the pattern.
It then searches the text for the lowlight character.
If there is a match, then compare other characters in the pattern string with corresponding characters
 in the text.  Usually, the method matches the pattern neither from the right nor
  from the left end, instead a special position.

  We now describe the algorithm as follows. Suppose that the text is $T_1\cdots T_n$, and the pattern is
  $P_1\cdots P_m$, where $n\geq m$.

\begin{itemize}
\item[(1)] \textit{Find a lowlight character in the pattern.}  If there are several characters in the pattern  which have the same smallest probability, pick the rightmost character. For example, $P_i$ is taken as the lowlight character.  Let the left segment be $P_1\cdots P_{i-1},$ and the right segment be $P_{i+1}\cdots P_m$.

\item[(2)] \textit{Search the text for the first mismatch}.

 \hspace*{8mm}\begin{minipage}{\linewidth}

   2-1.   Compare $T_i$ with $P_i$.  If $T_i\neq P_i$,  go to step 3-1.

    2-2.  Compare the left segment. Start the comparisons from the right end of the left segment, i.e., comparing $P_{i-\ell}$ with $T_{i-\ell}$, $\ell=1, \cdots, i-1$, one after another. Once there is a mismatch, go to step 3-2.

2-3.  Compare the right segment. Start the comparisons from the right end of the right segment, i.e., comparing $P_{i+\ell'}$ with $T_{i+\ell'}$, $\ell'=m-i, \cdots, 1$, one after another. Once there is a mismatch, go to step 3-3.

 \end{minipage}

\item[(3)] \textit{Align the pattern with the text}.

\hspace*{8mm}\begin{minipage}{\linewidth}

3-1. If  $T_i\neq P_{i-1},\cdots, P_1$,  then align $P_1$ with $T_{i+1}$. If $T_i\neq P_{i-1}, \cdots, P_{k+1},$ and $T_i=P_{k}, 1\leq k<i$, then align $P_k$ with $T_{i}$.

3-2.  Suppose that the mismatch appears at the position $s$, namely, $T_s\neq P_s$.  If $T_s\neq P_{s-1}, \cdots, P_1$, then align $P_1$ with $T_{s+1}$. If $T_s\neq P_{s-1}, \cdots, P_{l+1}$, and $T_s=P_l$, $1\leq  l<s$, then align $P_l$ with $T_s$.

 3-3 Suppose that the mismatch appears at the position $s'$, namely, $T_{s'}\neq P_{s'}$.  If $T_{s'}\neq P_{s'-1}, \cdots, P_{1}$, then align $P_1$ with $T_{s'+1}$. If $T_{s'}\neq P_{s'-1}, \cdots, P_{l'+1}$, and $T_{s'}=P_{l'}$, $1\leq  l'<s'$, then align $P_{l'}$ with $T_{s'}$.
\end{minipage}

\end{itemize}

Now we provide some examples to explain how to use the method.

\textbf{Example 1}:  Text is ``there they are".  Pattern is ``they".

Since $P_4$(\textit{y}) has the smallest probability 0.020 in the pattern, pick it as the lowlight character.
 Compare $P_4$(\textit{y}) with $T_4$(\textit{r}). It is a mismatch.

Since $T_4$(\textit{r}) does not appear in the pattern,  align $T_5$(\textit{e})  with $P_1$(\textit{t}).  Now compare  $P_4$(\textit{y}) with $T_8$(\textit{h}). It is a mismatch, too.

  Since $T_8$(\textit{h})=$P_2$(\textit{h}), align them and compare other characters.

  Like the Boyer-Moore algorithm, the new algorithm needs only 6 comparisons.

\textbf{Example 2}: Text is ``attach attack attain attempt attend
attention attest approve". Pattern is ``attempt".

Since $P_6$(\textit{p}) has the smallest probability 0.019 in the pattern, pick it as the lowlight character.
 Compare $P_6$(\textit{p}) with $T_6$(\textit{h}). It is a mismatch.

Since $T_6$(\textit{h}) does not appear in the pattern, align $T_7$(blank) with $P_1$(\textit{a}). Compare
$P_6$(\textit{p}) with $T_{12}$(\textit{c}).

Since $T_{12}$(\textit{c}) does not appear in the pattern, align $T_{13}$(k) with $P_1$(\textit{a}). Compare
$P_6$(\textit{p}) with $T_{18}$(\textit{a}).

Since $T_{18}$(\textit{a})=$P_1$(\textit{a}), align them. Compare
$P_6$(\textit{p}) with $T_{23}$(\textit{t}).

Since $T_{23}$(\textit{t}) =$P_3$(\textit{t}), not $P_2$(\textit{t}) (see the description of the new algorithm), align them. Compare
$P_6$(\textit{p}) with $T_{26}$(\textit{m}).

Since $T_{26}$(\textit{m}) =$P_5$(\textit{m}), align them. Compare
$P_6$(\textit{p}) with $T_{27}$(\textit{p}). Compare other characters.
Finally, we find the first appearance of the pattern in the text.

The new algorithm needs 12 comparisons. Note that the Boyer-Moore algorithm needs 10 comparisons for this example.

As mentioned earlier, the Boyer-Moore algorithm matches  from the right end of the pattern in order to move right more characters once a mismatch occurs. It is more appropriate for a text of plenty of words with
a same prefix. But it is insufficient for dealing with a text of plenty of words with
a same suffix which is just the suffix of the pattern.
Too see the shortcoming of the Boyer-Moore algorithm, we refer to the following example.

\textbf{Example 3}. The text is
``bear dear fear gear hear near pear rear sear tear wear year", and the pattern is ``wear".

In this example, the new algorithm picks \textit{w} as its lowlight character and matches from the left end of the pattern like the Knuth-Morris-Pratt algorithm. It saves much cost than the Boyer-Moore algorithm.

\subsection{Refined algorithm}

Note that there are two key factors for evaluating the new algorithm:
 \begin{itemize} 
 \item[(1)] the position for starting the comparisons in each shift is optimal;
  \item[(2)] the probability of the character in the position is small enough.\end{itemize}
 It seems difficult to balance exactly the two requirements. We suggest to take the following measure.

  \emph{Refined measure}:
  Find a lowlight character in the right half of the pattern, instead of the whole pattern. If the chosen character is at the right end of the pattern and it is a component of a common suffix, pick the  next-to-last position for starting comparisons.
   If there are several characters in the right half of the pattern  which have the same smallest probability, pick the rightmost character in the right half.

   Clearly, the refined measure has no effect on choosing the starting positions in the patterns ``they" and ``attempt" in example 1 and example 2, separately.  But it saves much cost when we use it to deal with the example 3.  In such case, the chosen character is $P_3$(\textit{a}). See the following process for details.

\begin{center}
   \begin{tabular}{|l|l|}
     \hline
    Text& be\textbf{a}r dear \,fear gear hear near pear rear sear tear wear year\\
 Pass 1& we\textbf{a}r \hspace*{8mm} --- 3 comparisons: a\&a; e\&e; w\&b  \\
   Text& bear \textbf{d}ear \,fear gear hear near pear rear sear tear wear year\\
 Pass 2& \ \ \ \ we\textbf{a}r  \hspace*{8mm} --- 1 comparison: a\&d\\
 Text& bear dea\textbf{r} \,fear gear hear near pear rear sear tear wear year\\
 Pass 3& \ \ \ \ \ \ \ \ we\textbf{a}r  \hspace*{8mm} --- 1 comparison: a\&r\\ 
  Text& bear dear \,f\textbf{e}ar gear hear near pear rear sear tear wear year\\
 Pass 4& \ \ \ \ \ \ \ \ \ \ \ \ we\textbf{a}r  \hspace*{8mm} --- 1 comparison: a\&e\\ \hline
  Text& bear dear \,fe\textbf{a}r gear hear near pear rear sear tear wear year\\
 Pass 5& \ \ \ \ \ \ \ \ \ \ \ \ \ \,we\textbf{a}r  \hspace*{8mm} --- 3 comparisons: a\&a; e\&e; w\&f\\ 
 \ \ \ $\vdots$ &  \hspace*{28mm} $\vdots$  \\
     \hline
   \end{tabular}\end{center}

\subsection{Complexity analysis}

Generally, it is reasonable to assume that
the pattern is not meaningless. Suppose that the chosen lowlight character \textit{P} in the pattern has the probability $\lambda$, the length of the text is $n$ and the length of the pattern is $m$. Hence, the text has about $\lambda n$ lowlight character \textit{P}. The amount of comparisons depends essentially  on the number of shifts. It is expected that the chosen lowlight character \textit{P} is at the $m/2$-th position in the pattern and the pattern moves right $m/4$ characters in each shift. Note that each shift is expected to has only one comparison because the first comparison happens to \emph{the starting character \textit{P}
in the pattern which is rarely matched}. Thus, the new algorithm needs about $4n/m$ comparisons. We conjecture that the character comparisons in the new algorithm is of order $\Theta(n/\lambda m)$ provided that $\lambda m \geq 1$.
As for the matching time analysis of the Knuth-Morris-Pratt algorithm and the Boyer-Moore algorithm, we refer to \cite{GG91,GG92}.

\begin{center}
\begin{tabular}{|l|l|}
  \hline
    Algorithm & Matching time \\ \hline
  Standard  algorithm & $\Theta((n-m+1)m)$  \\
   Knuth-Morris-Pratt algorithm  & $\Theta(n)$  \\
   Boyer-Moore algorithm &  $ \Omega(n/m), O(nm)$ \\
   New algorith & $\Theta(n/\lambda m), \lambda m\geq 1$ (it is conjectured)  \\
  \hline
\end{tabular}\end{center}

 \section{Conclusion }

 In this paper, we make use of the statistical probabilities of
English alphabets in natural language texts  to design a new
algorithm for string matching.
We hope the presentation could interest some skillful engineers to
experiment on it.

      \end{document}